\documentclass{jps-cp}
\usepackage{txfonts} %Please comment out this line unless the txfonts package is availabe in your LaTeX system.

\usepackage{dcolumn}
\usepackage{mathrsfs}
\usepackage{bm}
\usepackage{braket}
\usepackage{mathtools}
\usepackage{xcolor}

\title{Ground-State Properties of a Lattice-Coupled Strongly Correlated Model for Solid Oxygen under Pressure}

\author{
  Naomichi \textsc{Sato}$^{1}$,
  Makoto \textsc{Naka}$^{2}$,
  and Joji \textsc{Nasu}$^{1}$
}

\inst{
  $^{1}$Department of Physics, Graduate School of Science,
  Tohoku University, Sendai, Miyagi 980-8578, Japan \\
  $^{2}$School of Science and Engineering,
  Tokyo Denki University, Ishizaka 350-0394, Japan
}

\email{naomichi.sato.s7@dc.tohoku.ac.jp}

\recdate{\today}

\abst{
  Solid oxygen is an $S=1$ molecular magnet with strong
  coupling between spin and lattice degrees of freedom.
  Recent studies suggest that magnetic moments
  coexist with tetramer distortions under pressure.
  We derive an effective spin Hamiltonian from a two-orbital
  Peierls--Hubbard model with spin-orbit coupling.
  The resulting model contains competing easy-plane
  single-ion and easy-axis exchange anisotropies.
  We analyze this model on a square lattice using quantum Monte Carlo simulations
  and evaluate the magnetic contribution to the lattice stiffness 
  for several lattice distortion patterns including the tetramer pattern.
  We find that the tetramer pattern is the most favorable
  among the considered distortion patterns throughout the parameter range studied.
  Moreover, competition between the two anisotropies
  enhances the tendency toward lattice distortion
  and strengthens the preference for tetramerization over the other patterns.
}

\kword{solid oxygen, lattice distortion, spin Hamiltonian, quantum Monte Carlo}

\begin{document}
\maketitle

\section{Introduction}

The coupling of electronic degrees of freedom
to the lattice can stabilize diverse ordered phases
and generate unconventional physical responses.
This interplay is particularly important in molecular solids,
where weak intermolecular forces make the lattice soft
and allow molecular orientations and arrangements
to respond readily to changes in electronic states.
Solid oxygen provides a representative example,
with $S=1$ molecular magnetic moments arising from
two electrons in the doubly degenerate $\pi_g$ orbitals
of each $\mathrm O_2$ molecule.
The exchange interactions between these molecular magnetic moments
are closely coupled to the crystal structure.
This coupling contributes to the rich variety of
crystalline phases observed as temperature, pressure,
and magnetic field are varied
\cite{Freiman2004,Freiman2018,Nomura2014}.
At low temperatures, solid oxygen transforms
from the antiferromagnetic $\alpha$ phase through
the $\delta$ phase to the $\varepsilon$ phase
with increasing pressure \cite{Freiman2018}.
The $\varepsilon$ phase is characterized by the formation
of $(\mathrm O_2)_4$ tetramers and has traditionally
been regarded as nonmagnetic
\cite{Lundegaard2006,Fujihisa2006,Goncharenko2005}.
Recent single-crystal x-ray diffraction measurements
revealed an isostructural transition within the
$\varepsilon$ phase near $18$ GPa \cite{Gorelli2025}.
This transition was interpreted as structural evidence for
the collapse of molecular magnetic moments from
the lower-pressure $\varepsilon_1$ phase to
the higher-pressure $\varepsilon_0$ phase.
The survival of an $S=1$ magnetic moment on each
$\mathrm O_2$ molecule within a tetramer in the
$\varepsilon_1$ phase was previously proposed
by combining first-principles calculations with an
$S=1$ Heisenberg model \cite{Crespo2014}.
An earlier magnetoelastic analysis of an $S=1$
Heisenberg model coupled to lattice deformations
found that tetramer singlets have lower exchange
energy per molecule than dimer and trimer singlets
within an independent-cluster approximation
\cite{Gomonay2007}.
However, both spin-model analyses employ isotropic Heisenberg interactions 
and focus on cluster-singlet states.
To understand how magnetic interactions affect
the relative stability of competing lattice distortion
patterns in solid oxygen,
deviations from the isotropic spin model should be taken into account.

In this study, we investigate the magnetic mechanism
favoring tetramerization over competing lattice distortion
patterns in solid oxygen from a microscopic perspective.
We derive an effective $S=1$ spin Hamiltonian
from a two-orbital Peierls--Hubbard model
with spin-orbit coupling.
Using quantum Monte Carlo simulations of this model
on a square lattice, we evaluate the magnetic contribution to the lattice stiffness
from the quadratic coefficient of the free energy
to examine how the magnetic interactions affect
the stability of tetramerization.

\section{Model and Method}

\begin{figure}[tbp]
  \centering
  \includegraphics[width=1.0\textwidth,clip]{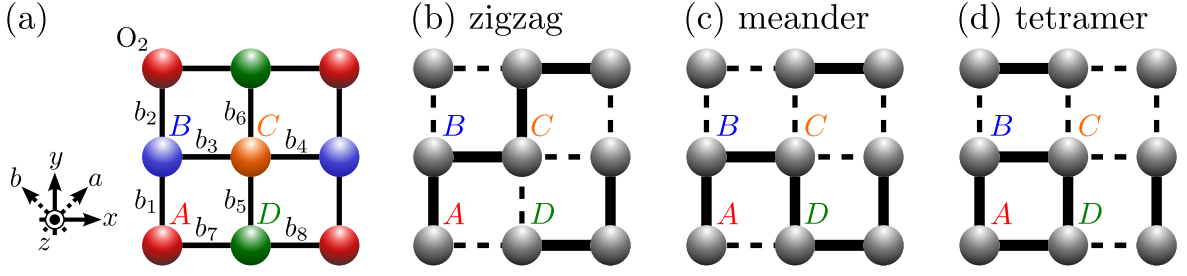}
  \caption{
    (a) Two-dimensional square lattice representing the molecular
    arrangement in the $ab$ plane of solid oxygen.
    Each site represents an $\mathrm O_2$ molecule.
    The four sublattices $A$, $B$, $C$, and $D$,
    the eight nearest-neighbor bond types $b_1,\ldots,b_8$,
    and the coordinate axes are indicated.
    (b) Zigzag, (c) meander, and (d) tetramer distortion patterns
    whose relative stability is examined in this study.
    Solid and dashed lines denote bond modulations
    $+\delta/2$ and $-\delta/2$, respectively.
    With the components ordered according to $b_1,\ldots,b_8$,
    the corresponding modulation vectors are
    $\bm v_{\rm Z}=(1,-1,1,-1,-1,1,-1,1)$,
    $\bm v_{\rm M}=(1,-1,1,-1,1,-1,-1,1)$, and
    $\bm v_{\rm T}=(1,-1,1,-1,1,-1,1,-1)$.
  }
  \label{fig:model_mond}
\end{figure}

To describe the coupling between electronic and lattice
degrees of freedom in solid oxygen, we employ a two-orbital
Peierls--Hubbard model on the square lattice shown in Fig.~\ref{fig:model_mond}(a).
Each site represents an $\mathrm O_2$ molecule with its axis
along $z$ and two electrons in the degenerate $\pi_g^x$ and $\pi_g^y$ orbitals.
The Hamiltonian is given by
$\mathcal H=\mathcal H_{U}+\mathcal H_{\rm SOC}+\mathcal H_{t}+\mathcal H_{K}$.
The onsite Coulomb interaction term is expressed as
$\mathcal H_U
=U\sum_{i,\mu}n_{i\mu\uparrow}n_{i\mu\downarrow}
+U'\sum_{i,\mu>\nu}n_{i\mu}n_{i\nu}
+J_{\rm H}\sum_{i,\mu>\nu}\sum_{ss'}
c^\dagger_{i\mu s}c^\dagger_{i\nu s'}
c_{i\mu s'}c_{i\nu s}$,
where $c^{\dagger}_{i\mu s}$ and $c_{i\mu s}$ are the creation and
annihilation operators of an electron with spin $s$ in orbital $\mu$ on
the $i$-th ${\rm O_2}$ molecule, respectively.
The electron number operators are defined as
$n_{i\mu s}=c^\dagger_{i\mu s}c_{i\mu s}$,
$n_{i\mu}=\sum_s n_{i\mu s}$, and
$n_i=\sum_\mu n_{i\mu}$,
where $n_i$ counts the total number of electrons at site $i$.
The orbital indices $\mu,\nu$ take the values $x$ and $y$,
corresponding to the $\pi_g^x$ and $\pi_g^y$ orbitals.
Here, $U$ and $U'$ are the intra- and inter-orbital Coulomb repulsions, and $J_{\rm H}$ is the Hund's coupling.
The spin-orbit coupling term with coupling
constant $\lambda$ is given by
$\mathcal H_{\rm SOC}
=\lambda\sum_{i,\mu\nu,ss'}
c^\dagger_{i\mu s}
(\ell^z)_{\mu\nu}(s^z)_{ss'}
c_{i\nu s'}$,
where
$\ell^z=\bigl(
\begin{smallmatrix}
0&-i\\
i&0
\end{smallmatrix}
\bigr)$
and $s^z=\sigma^z/2$.
Here, $\ell^z$ and $s^z$ denote the orbital and spin angular
momentum operators along the $z$ direction in units of $\hbar$,
and $\sigma^z$ denotes the corresponding Pauli matrix.
In the orbital basis $\{|\pi_g^x\rangle,|\pi_g^y\rangle\}$, the matrix elements of
$\ell^x$ and $\ell^y$ vanish, so the spin-orbit coupling reduces to
the $\ell^z s^z$ component.
The hopping term is expressed as
$\mathcal H_t
=-\sum_{\langle i,j\rangle}\sum_{\mu,\nu,s}
(1+\delta_{ij})
\left(
t^{ij}_{\mu\nu}c^\dagger_{i\mu s}c_{j\nu s}
+{\rm H.c.}
\right)$.
Here, $\langle i,j\rangle$ denotes a nearest-neighbor pair
on the square lattice, and $t^{ij}_{\mu\nu}$ is the hopping
amplitude between orbital $\mu$ at site $i$ and orbital $\nu$
at site $j$ in the undistorted lattice.
The dimensionless Peierls modulation is given by
$\delta_{ij}=\alpha{\bm e}_{ij}\cdot(\bm u_i-\bm u_j)$,
where $\alpha$ is the electron-lattice coupling constant,
$\bm u_i$ is the molecular displacement at site $i$,
and ${\bm e}_{ij}$ is the unit vector along the bond
from site $j$ to site $i$.
The elastic term is expressed as
$\mathcal H_K=
\frac{1}{2}\sum_{\langle i,j\rangle}
K_{ij}[{\bm e}_{ij}\cdot(\bm u_i-\bm u_j)]^2$,
where $K_{ij}$ is the elastic constant associated with the bond $\langle i,j \rangle$.

In the strong-coupling regime, we derive an effective
$S=1$ spin Hamiltonian by treating
$\mathcal H_0=\mathcal H_U+\mathcal H_{\rm SOC}$
as the unperturbed Hamiltonian and $\mathcal H_t$
as the perturbation.
At each site, we retain the three low-energy states
of the local Hamiltonian corresponding to
the spin triplet at $\lambda=0$.
Spin-orbit coupling splits them into a lower $S^z=0$ level
and an upper $S^z=\pm1$ doublet.
This splitting gives rise to easy-plane single-ion anisotropy
perpendicular to the molecular axis.
Using these three states as the local basis at each site,
we derive the intersite interactions by a canonical
transformation up to second order in $\mathcal H_t$
\cite{Nasu2016}.
The virtual intermediate states have electron occupations
$(n_i,n_j)=(1,3)$ or $(3,1)$
on a pair of sites $i$ and $j$.
For the hopping processes in $\mathcal H_t$,
$t_x$ and $t_y$ denote the intraorbital hopping amplitudes
for the $\pi_g^x$ and $\pi_g^y$ orbitals,
while $t_{xy}$ denotes the interorbital hopping amplitude.
In the resulting effective Hamiltonian $\mathcal H_{\rm eff}$,
the hopping-induced terms can be separated into
contributions proportional to
${\mathcal T}_0=t_x^2+2t_{xy}^2+t_y^2$ and ${\mathcal T}_1=t_{xy}^2-t_xt_y$.
Collecting the latter into $\Delta\mathcal H$, we write
$\mathcal H_{\rm eff}
=\mathcal H_{\rm spin}+\Delta\mathcal H+\mathcal H_K$.
For nearest-neighbor bonds, interorbital hopping vanishes
by symmetry, and one intraorbital hopping amplitude
is expected to dominate owing to the directional
overlap of the $\pi_g$ orbitals.
This leads to $|\mathcal T_1|\ll \mathcal T_0$,
allowing us to neglect $\Delta\mathcal H$.
Omitting an additive constant, the spin Hamiltonian
$\mathcal H_{\rm spin}$ is given by
\begin{align}
\mathcal H_{\rm spin}
={}&D_0\sum_i(S_i^z)^2
+\sum_{\langle i,j\rangle}
(1+\delta_{ij})^2
\Bigg[
J\,\bm S_i\cdot\bm S_j
+\frac{J_{z^2}}{\sqrt{3}}
\left(
2S_i^zS_j^z-S_i^xS_j^x-S_i^yS_j^y
\right)
\Bigg]
,
\label{eq:H_spin}
\end{align}
where $\bm S_i$ denotes the effective $S=1$ spin operator.
Up to second order in $\lambda$, the coefficients are given by
\begin{equation}
D_0
=
\frac{\lambda^2}{\Delta_S}
+O(\lambda^4),
\qquad
J=\frac{{\mathcal T}_0}{\Delta_J}
\left(
1+\frac{\lambda^2}{2\Delta_J^2}
-\frac{2\lambda^2}{3\Delta_S^2}
\right)
+O(\lambda^4),
\qquad
J_{z^2}=\frac{\lambda^2{\mathcal T}_0}
{\sqrt{3}\Delta_J\Delta_S^2}
+O(\lambda^4),
\label{eq:exchange_coefficients}
\end{equation}
where $\Delta_S=U-U'+J_{\rm H}$ and $\Delta_J=U+J_{\rm H}$.
The single-ion anisotropy $D_0>0$ and the uniaxial exchange
anisotropy $J_{z^2}>0$ compete in determining the preferred
spin orientation.
The former favors spins in the $xy$ plane,
whereas the latter favors antiferromagnetic alignment
along the $z$ axis.
Neutron diffraction measurements indicate that the spins
in solid oxygen lie approximately in the $ab$ plane
\cite{Meier1984,Klotz2010}.
This observation suggests that the easy-plane anisotropy
associated with $D_0$ dominates over the competing
uniaxial exchange anisotropy in our model.

%Using the effective spin Hamiltonian in Eq.~\eqref{eq:H_spin},
%we examine whether tetramerization is favored over competing
%lattice distortion patterns.
To accommodate the tetramerization observed in the
$\varepsilon$ phase of solid oxygen,
we use a four-sublattice unit cell with sublattices
$A$, $B$, $C$, and $D$ and eight independent nearest-neighbor
bonds $b_1,\ldots,b_8$, as shown in Fig.~\ref{fig:model_mond}(a).
Each nearest-neighbor bond belongs to one of these
eight types, and $\delta_{ij}$ takes the corresponding
value $\delta_\ell$ for $b_\ell$.
We consider the zigzag, meander, and tetramer distortion patterns
discussed in previous studies \cite{Aits2006}
and shown in Figs.~\ref{fig:model_mond}(b)--(d), respectively.
These patterns are also stabilized in our Hartree--Fock
analysis of a multiorbital Peierls--Hubbard model on a square lattice for solid oxygen
with self-consistently determined lattice displacements
\cite{Sato2026arXiv}.
In the following, these patterns are denoted by
$\eta=\mathrm{Z},\mathrm{M},\mathrm{T}$.
For each pattern, the eight bond modulations are collected
into the vector $\bm{\delta}_\eta=(\delta_1,\ldots,\delta_8)$
and parametrized as $\bm{\delta}_\eta=(\delta/2)\bm v_\eta$,
where $\bm v_\eta$ is an eight-component vector
with entries $\pm1$.
The amplitude $\delta$ represents the modulation contrast
within each bond pair $(b_1,b_2)$, $(b_3,b_4)$,
$(b_5,b_6)$, and $(b_7,b_8)$,
while the average modulation of each pair remains zero.
The explicit vectors are given in the caption of Fig.~\ref{fig:model_mond}.
We write $\mathcal H_{\rm spin}(\bm{\delta}_\eta)$
for the Hamiltonian in Eq.~\eqref{eq:H_spin}
with the bond modulations specified by $\bm{\delta}_\eta$.
Following Ref.~\cite{Aits2006}, we expand the total free energy
for each distortion pattern as
$F_\eta(\delta)=F^0+\frac{1}{2}N_{\rm site}
(K+a_\eta)\delta^2+O(\delta^4)$,
where $F^0$ is the free energy of the undistorted system.
The elastic and magnetic contributions to the lattice stiffness
per site are denoted by $K$ and $a_\eta$, 
with $K>0$ assumed to be common to all patterns.
Since the elastic term opposes lattice distortion,
a more negative $a_\eta$ enhances the tendency toward distortion.
The magnetic contribution $a_\eta$ is evaluated as
\begin{equation}
a_\eta
=
\frac{1}{N_{\rm site}}
\left\langle
\frac{\partial^2\mathcal H_{\rm spin}(\bm{\delta}_\eta)}
{\partial\delta^2}
\right\rangle_0
-
\frac{1}{N_{\rm site}}
\int_0^\beta d\tau
\left[
\langle D_\eta(\tau)D_\eta(0)\rangle_0
-\langle D_\eta\rangle_0^2
\right],
\label{eq:curvature}
\end{equation}
where $N_{\rm site}$ is the number of sites,
$D_\eta=
\left.\partial\mathcal H_{\rm spin}(\bm{\delta}_\eta)/
\partial\delta\right|_{\delta=0}$
is the bond-order operator associated with
the distortion pattern $\eta$,
$\beta=1/(k_{\rm B}T)$ is the inverse temperature,
and $\langle\cdots\rangle_0$ denotes a thermal expectation value
in the undistorted system.
The imaginary-time dependence is defined by
$D_\eta(\tau)=
\exp\{\tau\mathcal H_{\rm spin}(\bm 0)\}
D_\eta
\exp\{-\tau\mathcal H_{\rm spin}(\bm 0)\}$.
We introduce $a_\eta^{\rm bare}$ and $a_\eta^{\rm response}$ 
as the first and second terms in Eq.~\eqref{eq:curvature}, respectively.
The explicit quadratic dependence of the spin Hamiltonian
on $\delta$ gives rise to $a_\eta^{\rm bare}$,
whereas bond-order fluctuations associated with the
distortion pattern give rise to $a_\eta^{\rm response}$.

To evaluate these quantities, we simulate the spin Hamiltonian
using the stochastic series expansion quantum Monte Carlo method
\cite{Sandvik1999SSE} implemented in the Julia package
\texttt{Stochastic\allowbreak{}Series\allowbreak{}Expansion.jl} \cite{Weber2024SSE}.
We use a square-lattice cluster with $N_{\rm site}=128$ sites
and periodic boundary conditions.
The inverse temperature is fixed at $\beta J=10$.
For each parameter set, we perform ten independent runs
with different random-number seeds.
Measurements are binned within each run
to account for autocorrelations.
The reported values are averages over the ten runs,
and error bars represent one standard error of the mean
estimated from the variation among the run averages.

\section{Results and Discussion}
We compare the stability of the three distortion patterns
by evaluating their magnetic contributions $a_\eta$
to the lattice stiffness using Eq.~\eqref{eq:curvature}.
We take the antiferromagnetic isotropic exchange $J>0$
as the energy unit and vary $D_0/J$ and $J_{z^2}/J$
independently.
The calculations are performed for $D_0/J=0.0$, $0.2$, and $0.4$
over the range $0\leq J_{z^2}/J<0.15$.
Motivated by solid oxygen, we focus on positive anisotropy
parameters, for which the easy-plane single-ion anisotropy
competes with the easy-axis exchange anisotropy.
The case $D_0=0$ is included for comparison.

\begin{figure}[tbp]
  \centering
  \includegraphics[width=1.0\textwidth,clip]{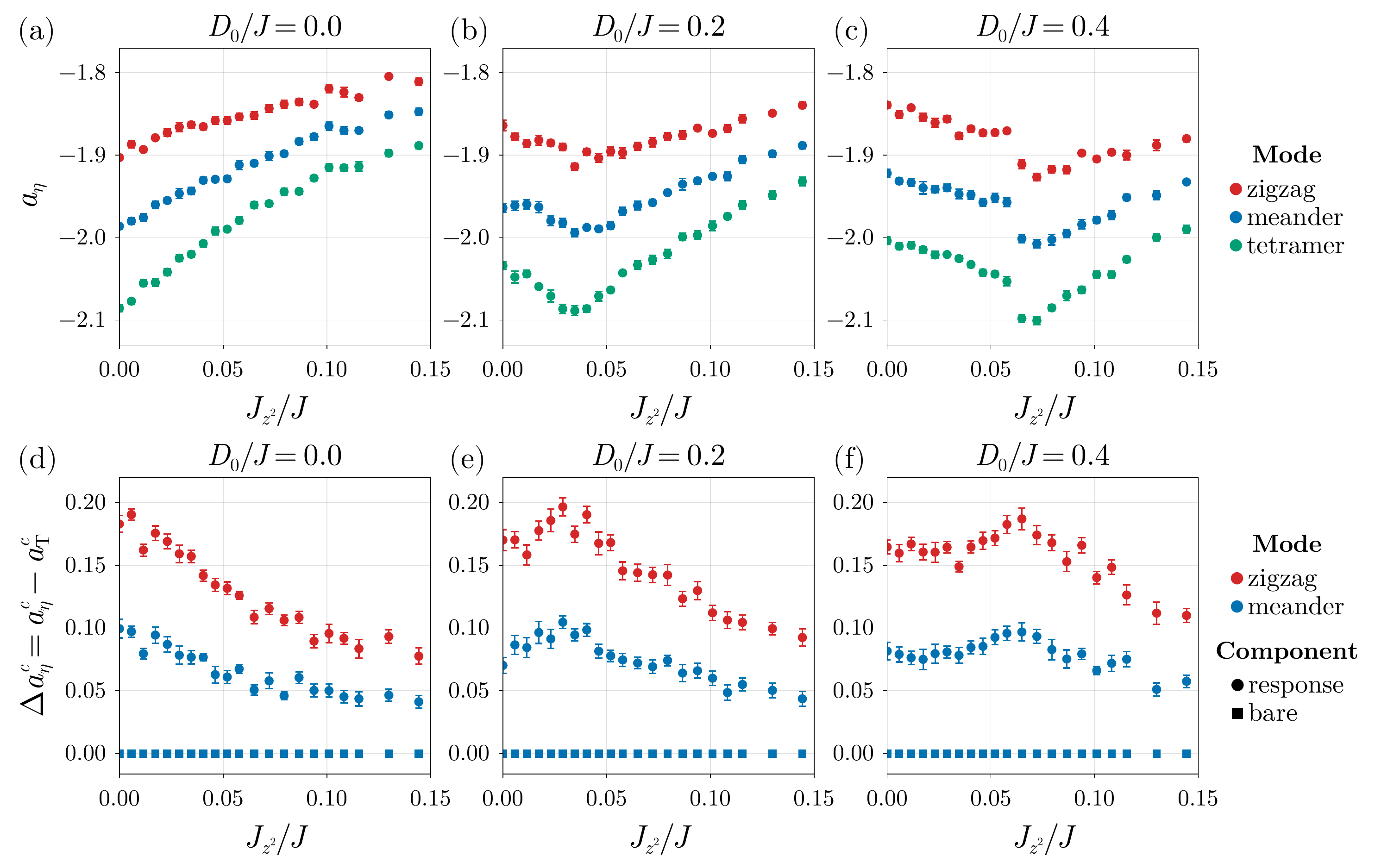}
  \caption{
    (a)--(c) Magnetic contributions $a_\eta$ to the lattice
    stiffness per site as functions of $J_{z^2}/J$ for
    $D_0/J=0.0$, $0.2$, and $0.4$, respectively.
    The symbols represent the zigzag (red), meander (blue), and tetramer (green) patterns.
    (d)--(f) Bare term $\Delta a_\eta^{\rm bare}$
    and response term $\Delta a_\eta^{\rm response}$
    in the differences relative to the tetramer pattern
    for the same values of $D_0/J$ as in (a)--(c).
    Circles and squares denote the response and bare
    contributions, respectively.
    The bare contributions overlap at zero.
    All values and their differences are expressed in units of $J$.
  }
  \label{fig:curvature}
\end{figure}

Figures~\ref{fig:curvature}(a)--(c) show the magnetic
contributions $a_\eta$ to the lattice stiffness
for the zigzag, meander, and tetramer patterns
as functions of $J_{z^2}/J$ for the three values of $D_0/J$.
All three values are negative throughout the parameter
range studied, with the tetramer pattern consistently
having the most negative $a_\eta$.
Thus, among the three patterns, tetramerization produces
the largest reduction in the spin free energy
for the same small distortion amplitude.
This preference is consistent with the stabilization
of the plaquette pattern found in the $S=1/2$
square-lattice Heisenberg model \cite{Aits2006}.
The dependence on the exchange anisotropy differs
between $D_0=0$ and $D_0>0$.
For $D_0=0$, $a_\eta$ becomes less negative
as $J_{z^2}/J$ increases, indicating a reduced tendency
toward lattice distortion.
For $D_0>0$, by contrast, $a_\eta$ initially decreases
and reaches a minimum at finite $J_{z^2}/J$ before increasing.
The minima shift to larger $J_{z^2}/J$ as $D_0/J$ increases.

To quantify the preference for tetramerization, we define
$\Delta a_\eta=a_\eta-a_{\mathrm T}$
for $\eta=\mathrm Z,\mathrm M$.
The contributions
$\Delta a_\eta^c=a_\eta^c-a_{\mathrm T}^c$
with $c=\mathrm{bare},\mathrm{response}$
are shown in Figs.~\ref{fig:curvature}(d)--(f).
The bare term is identical for all three patterns
because the second derivative of each exchange coefficient
depends on the square of the corresponding component
of $\bm v_\eta$, which is unity.
Consequently, $\Delta a_\eta^{\rm bare}=0$,
and the preference for tetramerization is entirely determined by the response term.
For $D_0=0$, the positive differences $\Delta a_\eta$ decrease
overall with increasing $J_{z^2}/J$,
so the relative advantage of tetramerization decreases.
For $D_0>0$, the differences instead reach maxima
at finite $J_{z^2}/J$ and decrease at larger values.
Increasing $J_{z^2}/J$ up to these maxima therefore
further stabilizes the tetramer pattern relative to
the zigzag and meander patterns.

\begin{figure}[tbp]
  \centering
  \includegraphics[width=1.0\textwidth,clip]{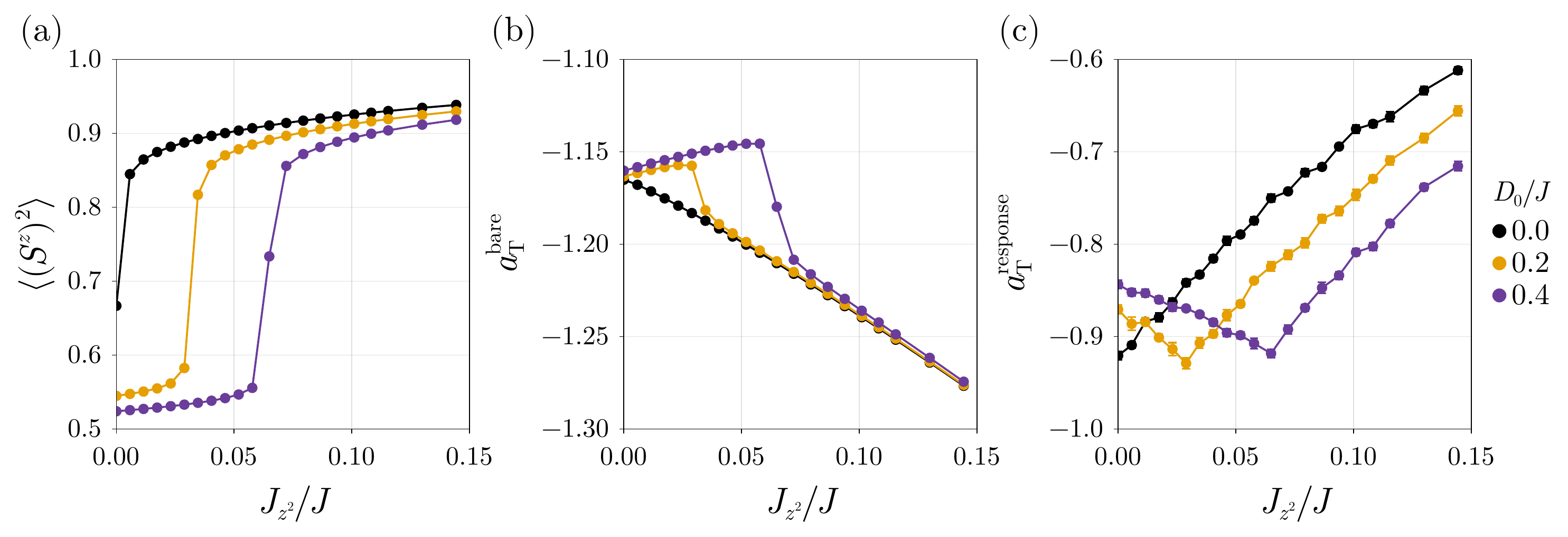}
  \caption{
    (a) Mean-square $z$ component of the spin, $\langle(S^z)^2\rangle$.
    (b) Bare and (c) response terms in the magnetic contribution 
    to the lattice stiffness per site for the tetramer pattern.
    These terms are expressed in units of $J$.
    All quantities are evaluated in the undistorted system
    and plotted as functions of $J_{z^2}/J$.
    Black, orange, and purple symbols correspond to
    $D_0/J=0.0$, $0.2$, and $0.4$, respectively.
  }
  \label{fig:observables}
\end{figure}

These trends can be related to the evolution
of the local spin anisotropy.
Figure~\ref{fig:observables}(a) shows
the mean-square $z$ component of the spin,
$\langle(S^z)^2\rangle
=N_{\rm site}^{-1}\sum_i\langle(S_i^z)^2\rangle_0$,
in the undistorted system.
For $D_0>0$, this quantity remains close to $1/2$
at small $J_{z^2}/J$, consistent with an easy-plane tendency.
It rises sharply as $J_{z^2}/J$ increases further,
reflecting a change toward an easy-axis tendency.
This change occurs at larger $J_{z^2}/J$ as $D_0/J$ increases.
The minima in $a_\eta$ and the maxima in $\Delta a_\eta$
both occur near this change in spin anisotropy.
This correspondence suggests that competition between
the easy-plane single-ion anisotropy and the easy-axis
exchange anisotropy enhances both the tendency toward
lattice distortion and the relative stability of the
tetramer pattern.
At larger $J_{z^2}/J$, both effects weaken as the
easy-axis tendency becomes dominant.

To clarify how $a_{\mathrm T}$ depends on the exchange
anisotropy, we examine its bare and response terms
shown in Figs.~\ref{fig:observables}(b) and (c).
For $D_0>0$ in the region with an easy-plane tendency,
the bare term increases slightly toward zero
as $J_{z^2}/J$ increases, whereas the response term
decreases to more negative values.
The initial decrease in $a_{\mathrm T}$ is therefore
driven by the enhanced bond-order response.
Near the change toward an easy-axis tendency,
the bare term also decreases rapidly,
further lowering $a_{\mathrm T}$.
At larger $J_{z^2}/J$, the two terms show the same trends
as for $D_0=0$.
In both cases, the bare term decreases with increasing
$J_{z^2}/J$, but the response term increases by a larger amount.
Consequently, $a_{\mathrm T}$ becomes less negative.
The increase in the response term is consistent with
the increasingly Ising-like character of the exchange interactions.

The results suggest that competition between
the two anisotropies enhances the tendency toward lattice distortion.
In particular, the enhanced bond-order response further
favors tetramerization over the zigzag and meander distortions.
Both effects are strongest near the change
in the preferred spin orientation.
This provides a possible mechanism favoring tetramerization
in the effective spin model for solid oxygen.

\section{Conclusion}
In summary, we derived an effective $S=1$ spin Hamiltonian
from a two-orbital Peierls--Hubbard model with spin-orbit
coupling to investigate lattice instabilities in solid oxygen.
The resulting model contains competing easy-plane single-ion
and easy-axis exchange anisotropies.
The magnetic contributions to the lattice stiffness
obtained from quantum Monte Carlo simulations show that
tetramerization is favored over the zigzag and meander
distortions throughout the parameter range studied.
Competition between the two anisotropies further enhances
the relative stability of the tetramer pattern.
These findings show that magnetic interactions influence
the tendency toward lattice distortion and suggest that
competing magnetic anisotropies provide a possible mechanism
favoring tetramerization in solid oxygen.
The effects of the quadrupolar interactions neglected
in the present analysis remain to be investigated.
To describe the molecular-moment collapse associated
with the proposed $\varepsilon_1$--$\varepsilon_0$
transition, the effective model should also be extended
to include local $S=0$ spin-singlet states
alongside the triplet-based states.

\section*{Acknowledgments}
The authors thank T. Aoyama and T. Kimura for fruitful discussions.
This work was supported by Grants-in-Aid for Scientific Research from JSPS,
KAKENHI Grant Nos. JP20H00122, JP22H01175, JP23K25826, JP23H04865, JP24K00563, and
JP26H00624.

\bibliographystyle{sces-refs}
\bibliography{refs}

\end{document}